\documentclass[pra,aps,amssymb,amsmath,amsmath,showpacs,reprint,twocolumn]{revtex4-1}
\usepackage{color}
\usepackage{graphicx}
\usepackage{epstopdf}
\usepackage{natbib}
\usepackage{hyperref}
\usepackage{dcolumn}
\usepackage{bm}
\usepackage{dcolumn}
\usepackage{multirow}

\newcolumntype{d}[1]{D{.}{.}{#1}}

\newcommand{\Fkt}[1]{\,\mathsf {#1}}
\def\openone{\leavevmode\hbox{\small1\kern-3.3pt\normalsize1}}

\ifx\Tr\renewcommand{\Tr}{\Fkt{Tr}} 
\else\newcommand{\Tr}{\Fkt{Tr}}
\fi

\usepackage{booktabs}
\usepackage{graphicx}
\usepackage{amsmath}
\usepackage{indentfirst}
\hypersetup{colorlinks=true,citecolor=blue,linkcolor=blue,urlcolor=blue}
\begin{document}
\title{Relativistic treatment of diamagnetic susceptibility of helium}

\author{\sc Mariusz Puchalski}
\affiliation{Faculty of Chemistry, Adam Mickiewicz University, Uniwersytetu Pozna\'{n}skiego 8, 61-614 Pozna\'{n}, Poland}
\author{\sc Micha\l\ Lesiuk}
\email{e-mail: m.lesiuk@uw.edu.pl}
\author{\sc Bogumi\l\ Jeziorski}
\affiliation{\sl Faculty of Chemistry, University of Warsaw\\
Pasteura 1, 02-093 Warsaw, Poland}
\date{\today}

\begin{abstract}
 We report theoretical calculations of the diamagnetic susceptibility, $\chi_0$, of helium atom. 
We determined the complete relativistic correction to $\chi_0$ of the order of $\alpha^4$, where $\alpha$ is the fine structure constant, by including all $\alpha^4$ terms originating from the Dirac and Breit equations for a helium atom in a static magnetic field. Finite-nuclear-mass corrections to $\chi_0$ was also evaluated. To obtain very accurate results and reliable uncertainty estimates we used a sequence of explicitly correlated basis sets of fully optimized Slater geminals. We found that $\chi_0=-2.119\,070(34)\cdot10^{-5}$ $a_0^3$ and $\chi_0=-2.119\,365(34)\cdot10^{-5}$ $a_0^3$ for $^4$He and $^3$He isotopes, respectively, where $a_0$ is the Bohr radius and the uncertainties shown in the parentheses are due entirely to the very conservative estimate of the neglected QED corrections of the order of $\alpha^5$. Our results are compared with the available experimental data and with previous, incomplete theoretical determinations of the $\alpha^4$ contributions to the diamagnetic susceptibility of helium.
\end{abstract}

\maketitle

\section{Introduction}
\label{sec:intro}

For closed-shell atoms the static diamagnetic susceptibility $\chi_0 $ can be defined as the second derivative of the energy, $E$, with respect to the strength $B=|\mathbf{B}|$ of the uniform external magnetic field $\mathbf{B}$, in the limit of $B\rightarrow 0$,
\begin{align}
 \label{chi0}
 \chi_0 = - \frac{\partial^2 E}{\partial B^2}\bigg|_{B=0}.
\end{align}
In general, the magnetic susceptibility is dependent on the frequency of the oscillating magnetic field. However, as for closed-shall atoms the frequency-dependent terms appear only in the order of $\alpha^5$~\cite{yerokhin11,yerokhin12} or are quadratic in the electron-to-nucleus mass ratio~\cite{lesiuk20} and, consequently, very small, we consider only static magnetic fields. Our interest in this quantity is motivated primarily by recent advances in metrology~\cite{jousten17,gaiser18,gaiser20,gaiser22}. In particular, in the refractive-index gas thermometry~\cite{gao17,rourke19,ripa21,rourke21}, 
measurements of the refractive index $n$ of a gas are used to determine its density $\rho$. If the equation of state is known, for instance in the form of the virial expansion, the measurement of $n$ provides a possibility to find the gas pressure $p$. Alternatively, the thermodynamic temperature of a gas can be determined knowing its refractive index and pressure. The fundamental relation linking the refractive index and the gas density is the Lorentz-Lorenz formula~\cite{lorentz80,lorenz80}
\begin{align}
 \frac{n^2-1}{n^2+2} = \frac{4\pi}{3}\,\big(\alpha_{\rm d} +\chi\big)\rho,
\end{align}
where $\alpha_{\rm d}$ is the electric dipole polarizability of the gas particles. Formally, this expression is valid only for 
small densities, but generalizations involving higher powers of $\rho$ with extended range of applicability are 
well-known~\cite{jousten17}. 

In most realizations of the refractive-index gas thermometry, helium is used as a medium gas~\cite{rourke19}. Currently, the most reliable sources of fundamental microscopic properties of helium are \emph{ab initio} calculations. For example, the electric dipole polarizability, $\alpha_{\rm d}$, is known from theory with relative accuracy of
$10^{-7}$~\cite{pachucki00,lach04,Piszczatowski:2015,Puchalski:2016,puchalski20}. In the foreseeable future, we expect the present accuracy level for $\alpha_{\rm d}$ to be entirely sufficient from the experimental point of view. However, the same cannot be said about the magnetic susceptibility. On one hand, this quantity is roughly five orders of magnitude smaller than $\alpha_{\rm d}$ and hence does not have to be determined as accurately. On the other hand, the most recent calculations of $\chi_0$ by Bruch and Weinhold~\cite{Bruch:2002,Bruch:2003} for helium differ from the experimental results of Barter~et~al.~\cite{barter60} by roughly $7\%$. While the current consensus is that such large discrepancy is most likely due to errors in the measurements, some problems on the theoretical side remain. As pointed out by Pachucki~\cite{pachucki03}, the relativistic correction to $\chi_0$ calculated by Bruch and Weinhold is incomplete and misses several terms originating from the magnetic-field dependence of the Dirac equation and the Breit interaction. As the magnitude of these terms is yet 
unknown, it is impossible to rigorously determine the uncertainty of the calculated $\chi_0$. 

This situation is not satisfactory form the point of view of modern metrological applications. Additionally, 
refractive-index gas thermometry measurements using neon and argon as medium gas have been proposed and argued to offer 
several advantages over helium~\cite{rourke21}. Unfortunately, the magnetic susceptibility of neon and, especially, 
argon is not known with sufficient accuracy. This prompted us to undertake a systematic theoretical calculations of the 
static magnetic susceptibility of light noble gases: helium, neon and argon. Our results for helium-4 and helium-3 are reported in the present paper, while the magnetic susceptibility of neon and argon are considered in the subsequent publication.

Throughout most this work, we use the standard c.g.s. system of units employed, for instance, in the book of Bethe and Salpeter~\cite{Bethe:77}. In these units, employed also in the experimental work, the magnetic susceptibility has the dimension of the volume and is expressed in cm$^3$/mol. 
In order to present the intermediate and final results of our calculations it is convenient to use the atomic units (a.u.), where the electron mass $m_e$, charge $e$ and the Planck constant $\hbar$ are assumed to be equal to 1. The atomic unit of the magnetic susceptibility is then $a_0^3$, where $a_0$ is the Bohr radius, $a_0 = 0.529\,177\,211 \times 10^{-8}$ cm. The conversion relation between the c.g.s and the atomic unit is cm$^3$/mol = 11.205\,873\,1~$a^3_0$. For the speed of light in vacuum we adopt the value $c=\alpha^{-1}=137.035\,999$ atomic units. The masses of the alpha particle and the helium-3 nucleus used by us are $7\,294.299\,54$ and $5\,495.885\,28$ a.u., respectively.

\section{Theory}
\label{sec:theory}

\subsection{Leading-order contribution}

Consider a helium atom in its electronic ground, $^1S$ state. We temporarily neglect the nuclear motion and treat the nucleus as a stationary classical charge with infinite mass. Let us denote the non-relativistic electronic Hamiltonian in the absence of external fields by $\hat{H}_0$. The total Hamiltonian $\hat{H}$ in the magnetic field 
$\mathbf{B}$ reads then~\cite{Bethe:77} 
\begin{align}
\begin{split}
 \hat{H} &= \hat{H}_0 
 + \frac{e}{mc}\,\mathbf{B}\cdot \mathbf{S} 
 + \frac{e}{mc}\,\mathbf{B}\cdot \mathbf{L} \\ \label{Hnonrel}
 &+ \frac{e^2}{8mc^2} \left[ 
 \left( \mathbf{B}\times\mathbf{r}_1 \right)^2 
 + \left( \mathbf{B}\times\mathbf{r}_2 \right)^2 
 \right],
\end{split}
\end{align}
where $\mathbf{S}$ and $\mathbf{L}$ are the total spin and angular momentum operators, respectively. The origin of the coordinate system is placed at the atomic nucleus and $\mathbf{r}_i$, $i=1,2$, denotes coordinates of $i$th electron with respect to the origin. In the non-relativistic theory the terms linear in $\mathbf{B}$ bring no contribution to the magnetic susceptibility of a $^1S$ state. The quadratic, diamagnetic term gives~\cite{Bethe:77,landau81}
\begin{align}
\chi_0^{(0)} = -\frac{e^2}{6mc^2}\,\langle \psi_0|r_1^2 + r_2^2|\psi_0 \rangle, 
\end{align}
where $\psi_0$ is the ground-state wave function. This is the dominant contribution to the magnetic susceptibility of helium. The leading corrections to $\chi_0^{(0)}$ computed in this work are either of the order $\alpha^4$ (referred to as the relativistic corrections) or are proportional the electron-to-nucleus mass ratio.

\subsection{Finite-nuclear-mass corrections}
\label{sec:fnm}

Finite-nuclear-mass (FNM) corrections to the magnetic susceptibility of helium were derived by Bruch and 
Weinhold~\cite{Bruch:2002}, see also the Erratum correcting a small numerical error~\cite{Bruch:2003}. For the helium atom at rest, the complete correction of the order $m_{\rm e}/m_{\rm N}$, where $m_{\rm N}$ is the mass of the 
nucleus, comprises three contributions, 
\begin{align}
 \chi_0^{\mathrm{FNM}} = \delta\chi_0^{\mathrm{ms }} + 
 \delta\chi_0^{\mathrm{PZW }} + 
 \delta\chi_0^{\mathrm{mp}}.
\end{align}
The first is the reduced-mass scaling term~\cite{ Bruch:2002}
 \begin{align} \label{ms}
 &\delta\chi_0^{\mathrm{ms}} = 3\frac{m}{m_N}\,\chi_0^{(0)} 
 \end{align}
 while the second results from the application of the Power-Zienau-Wooley transformation to eliminate the dependence of the vector potential on the center-of-mass position~\cite{Bruch:2002,Pachucki:2019}
 \begin{align} \label{PWZ}
 &\delta\chi_0^{\mathrm{PZW}} = -\frac{e^2}{3m_Nc^2}\,
 \langle\psi_0|\mathbf{r}_1\cdot \mathbf{r}_2| \psi_0\rangle 
\end{align} 
The third term is the correction due to the conventional mass-polarization term in the Hamiltonian resulting from the separation of the centre of mass motion, 

\begin{align}
\label{fnmmp}
 \delta\chi_0^{\mathrm{mp}} = -\frac{e^2}{3mc^2}\,
 \langle\psi_0|\big(r_1^2 + r_2^2\big)\mathcal{R}_0\,H_{\mathrm{mp}}|\psi_0\rangle, 
\end{align}
where $\mathcal{R}_0 = \big(E_0-Q\hat{H}_0\big)^{-1}Q$ with $Q=1-|\psi_0\rangle\langle\psi_0|$ being the ground-state 
reduced resolvent of $H_0$, and $H_{\mathrm{mp}}= {\mathbf{p}_1\cdot \mathbf{p}_2}/{m_N}$ is the 
mass-polarization perturbation. Equations (\ref{ms}) and (\ref{PWZ}) are special cases of the equations derived by Pachucki and Yerokhin \cite{Pachucki:2019} for many electron atoms. Bruch and Weinhold \cite{Bruch:2002} considered also a small temperature-dependent correction,
denoted as $\delta\chi_0^{\mathrm{BO}}$, 
resulting from the center-of-mass motion of the atom and derived an order-of-magnitude estimation of its value. The significance of this correction will be discussed in Sec. IV. 

\subsection{Relativistic corrections}
\label{sec:rel}

Relativistic corrections to the magnetic susceptibility can be divided into three groups. 
The first group originates from the Foldy-Wouthuysen transformation of the Dirac Hamiltonian in the presence of homogeneous external magnetic field. The transformed Hamiltonian contains several 
magnetic-field-dependent terms, see Eq.~(14) in Ref.~\cite{Pachucki:2008}, that are not included in Eq. (\ref{Hnonrel}). There are two terms linear in the magnetic-field vector $\mathbf{B}$, which can give a contribution of the order of $\alpha^6$ (and also a small frequency dependence of $\chi_0$) 
and, therefore, are beyond the scope of the present work. Eq.~(14) of Ref.~\cite{Pachucki:2008} contains also four diamagnetic terms quadratic in $\mathbf{B}$ that read
\begin{align}
 \hat{A}^{(0)} &= \frac{e^2}{4m^3c^4}\Big[ 
 \big\{ \mathbf{l}_1\cdot\mathbf{B},\,\mathbf{s}_1\cdot\mathbf{B} \big\} +
 \big\{ \mathbf{l}_2\cdot\mathbf{B},\,\mathbf{s}_2\cdot\mathbf{B} \big\} \Big],\\
 \hat{A}^{(1)} &= -\frac{e^2}{8m^3c^4}\big[ \big( \mathbf{l}_1\cdot\mathbf{B} \big)^2
 + \big( \mathbf{l}_2\cdot\mathbf{B} \big)^2\big], \\
 \begin{split}
 \hat{A}^{(2)} &= -\frac{e^2}{32m^3c^4}\Big[ 
 \big\{ (\mathbf{B}\times\mathbf{r}_1)^2, \mathbf{p}_1^2 \big\} \\
 &+ \big\{ (\mathbf{B}\times\mathbf{r}_2)^2, \mathbf{p}_2^2 \big\} \Big],
 \end{split}\\
 \hat{A}^{(3)} &= -\frac{e^2 \hbar^2}{4m^3c^4}B^2, 
\end{align}
where $\mathbf{p}_i$ and $\mathbf{l}_i = \mathbf{r}_i\times \mathbf{p}_i$ are the momentum and angular momentum operators, respectively, of the $i$-th electron. The curly brackets in the above formulas denote the anti-commutators. Note the additional factor of $2$ included in Eq.~(12) compared to the value given in Eq.~(14) of Ref.~\cite{Pachucki:2008}. This change is due to the fact that Eq.~(14) in Ref.~\cite{Pachucki:2008} applies to one electron only while we are considering a two-electron system. For singlet states, $\hat{A}^{(0)}$ gives no $\alpha^4$ contribution to $\chi_0$, since its expectation value vanishes under spin integration. By differentiating the expectation values of the operators $ \hat{A}^{(1)}$, $ \hat{A}^{(2)}$, and $ \hat{A}^{(3)}$ with respect to $ {B}$ and setting the magnetic-field strength equal to zero, one obtains consecutively three corrections to the static magnetic susceptibility
\begin{align}
\label{chi1}
 &\delta\chi_0^{(1)} = \frac{e^2}{12m^3c^4}\,\langle\psi_0|l_1^2 + l_2^2|\psi_0\rangle, \\ \label{chi2}
 &\delta\chi_0^{(2)} = \frac{e^2}{12m^3c^4}\,
 \langle\psi_0|r_1^2p_1^2 + r_2^2p_2^2|\psi_0\rangle, \\ \label{chi3}
 &\delta\chi_0^{(3)} = \frac{e^2\hbar^2}{2m^3c^4}. 
\end{align}

The second group of relativistic contributions originates from the Breit correction to the 
electron-electron interaction. The explicit form of the Breit-Pauli Hamiltonian in the presence of homogeneous electric and magnetic 
fields has been given in Ref.~\cite{Pachucki:2008}, see Eq.~(17) of this reference. Similarly as for the Dirac Hamiltonian, it contains several linear and quadratic magnetic-field dependent terms. However, all terms linear in $\mathbf{B}$ give contributions to 
$\chi_0$ that are of the order of $1/c^6$ and hence are neglected in the present work. The spin-dependent quadratic terms vanish for 
singlet states upon the spin integration. All spin-independent quadratic terms in the Breit-Pauli Hamiltonian, which can give an $\alpha^4$ contribution to 
$\chi_0$, originate from the orbit-orbit interaction
\begin{align} \label{oo}
 \hat{H}_{\mathrm{oo}} = -\frac{e^2}{2m^2c^2}\bigg[\frac{\bm{\pi}_1\cdot\bm{\pi}_2}{r_{12}}
 + \frac{\big(\bm{\pi}_1\cdot\mathbf{r}_{12}\big)\big(\mathbf{r}_{12}\cdot\bm{\pi}_2\big)}{r_{12}^3}
 \bigg],
\end{align}
where $\mathbf{r}_{12}=\mathbf{r}_1-\mathbf{r}_2$, and $\bm{\pi}_i = \mathbf{p}_1+e\,(\mathbf{B}\times\mathbf{r}_i)/2c$. From Eq.~(\ref{oo}) we obtain two diamagnetic terms quadratic in the magnetic-field vector, namely
\begin{align}
\label{A44}
 \hat{A}^{(4)} &= -\frac{e^4}{8m^2c^4}\,
 \frac{(\mathbf{B}\times\mathbf{r}_1)\cdot(\mathbf{B}\times\mathbf{r}_2)}{r_{12}}, \\
 \label{d6}
 \hat{A}^{(5)} &= -\frac{e^4}{8m^2c^4}\,
 \frac{\big[(\mathbf{B}\times\mathbf{r}_1)\cdot\mathbf{r}_{12}\big]
 \big[(\mathbf{B}\times\mathbf{r}_2)\cdot\mathbf{r}_{12}\big]}{r_{12}^3}.
\end{align}
Differentiation with respect to $B$ leads to the following two corrections: 
\begin{align}
\label{chi4}
 &\delta\chi_0^{(4)} = \frac{e^4}{6m^2c^4}\,
 \langle\psi_0|\frac{\mathbf{r}_1\cdot \mathbf{r}_2}{r_{12}}|\psi_0\rangle, \\
\label{chi5}
 &\delta\chi_0^{(5)} = \frac{e^4}{12m^2c^4}\,
 \langle\psi_0|\frac{\mathbf{r}_1\cdot \mathbf{r}_2}{r_{12}}
 -\frac{(\mathbf{r}_1\cdot\mathbf{r}_{12})(\mathbf{r}_2\cdot\mathbf{r}_{12})}{r_{12}^3}
 |\psi_0\rangle.
\end{align}
Derivation of Eqs.~(\Ref{chi4}) and (\Ref{chi5}) is given in Appendix~\ref{sec:app}.

Finally, the third group of contributions to the magnetic susceptibility originates from relativistic corrections to the electronic wavefunction. Using the standard perturbation theory one derives the following general formula
\begin{align}
\label{relbp}
 \delta\chi_0^{\mathrm{BP}} = -\frac{e^2}{3m c^2}\,
 \langle\psi_0|\big(r_1^2 + r_2^2\big)\mathcal{R}_0\,\hat{H}_{\mathrm{BP}}|\psi_0\rangle, 
\end{align}
where $\mathcal{R}_0$ is the resolvent defined in the same way as in Eq.~(\ref{fnmmp}), and $\hat{H}_{\mathrm{BP}}$ is the relativistic part of the Breit-Pauli Hamiltonian in the absence of the external electric and magnetic fields. When acting on singlet states 
this Hamiltonian can be assumed to comprise the following four terms~\cite{Bethe:77} 
\begin{align} \label{BP}
 &\hat{H}_{\mathrm{BP}} = \hat{P}_4 + \hat{D}_1 + \hat{D}_2 + \hat{B}, \\ \label{mv}
 &\hat{P}_4 = -\frac{1}{8m^3c^2}\left( p_1^4 + p_2^4 \right), \\ \label{Darwin1}
 &\hat{D}_1 = \frac{\pi e^2\hbar^2}{m^2c^2}
 \big[ \delta(\mathbf{r}_1) + \delta(\mathbf{r}_2) \big], \\ \label{Darwin2}
 &\hat{D}_2 = \frac{\pi e^2\hbar^2}{m^2c^2}\,\delta(\mathbf{r}_{12}),\\ \label{oo2}
 &\hat{B} = -\frac{e^2}{2m^2c^2}\left[\frac{\mathbf{p}_1\cdot \mathbf{p}_2}{r_{12}}
 -\frac{(\mathbf{p}_1\cdot\mathbf{r}_{12})(\mathbf{p}_2\cdot\mathbf{r}_{12})}{r_{12}^3}\right],
\end{align}
where $\delta(\mathbf{r})$ is the three-dimensional Dirac distribution. The terms in 
Eqs.~(\ref{mv})-(\ref{oo2}) are usually referred to as, consecutively, the mass-velocity, 
one-electron Darwin, two-electron Darwin, orbit-orbit interaction (or Breit) operators. For further convenience, we split the $\delta\chi_0^{\mathrm{BP}}$ correction into the components related to the individual operators in Eqs.~(\ref{mv})-(\ref{oo2}), 
\begin{align}
 \delta\chi_0^{\mathrm{BP}} = \delta\chi_0^{\mathrm{P}_4} + \delta\chi_0^{\mathrm{D}_1}
 + \delta\chi_0^{\mathrm{D}_2} + \delta\chi_0^{\mathrm{B}}.
\end{align}
In summary, the total relativistic correction to the static diamagnetic susceptibility, evaluated in the present work, comprises nine terms
\begin{align}
\label{reltot}
 \delta\chi_0^{\mathrm{rel}} = \sum_{i=1}^5 \delta\chi_0^{(i)} + \delta\chi_0^{\mathrm{P}_4} + \delta\chi_0^{\mathrm{D}_1}
 + \delta\chi_0^{\mathrm{D}_2} + \delta\chi_0^{\mathrm{B}}.
\end{align}
In the work of Bruch and Weinhold \cite{Bruch:2002}, only the last four terms in 
Eq.~(\ref{reltot}) were considered and all the remaining ones were neglected.

\subsection{Quantum electrodynamics correction}
\label{sec:qed}

The leading corrections to $\chi_0$ which have not been considered thus far originate from quantum electrodynamics (QED). These corrections are of the order $\alpha^5$ (in fact of order of $\alpha^5\log\alpha$) and account of two physical phenomena: vacuum polarization and electron self-energy. 
The QED formulas for these corrections can be derived along a similar lines as for the nuclear magnetic shielding constants~\cite{yerokhin11,yerokhin12,Wehrli:2021} and implemented numerically for helium in a way largely parallel to that presented in Ref.~\cite{Wehrli:2022}. The resulting computations would inevitably be extremely complicated as they would require calculations of new forms the so-called Bethe logarithms including their magnetic-field dependence~\cite{Wehrli:2022}. This would represent
a massive computational task far beyond the scope of the present work. 

However, one can easily perform a crude, order-of-magnitude assessment of the QED effects and obtain a conservative estimate of the uncertainty of $\chi_0$ computed by us that can be useful in metrological applications. From the formal perturbation theory expressions one can naively expect that QED corrections should be by a factor of the order of $\alpha\log\alpha$ smaller than the relativistic corrections determined in the present work.
 
However, it would be overly optimistic to scale the total relativistic correction to $\chi_0$ by $\alpha \log\alpha$. In fact, there is a considerable cancellation between various relativistic contributions, making the total correction significantly smaller than the individual contributions. It is impossible to guarantee that a similar cancellation persists also for the QED corrections. In fact, in the case of QED corrections to the energy~\cite{pachucki06,piszcz09,cencek12,lesiuk15,lesiuk19,Yerokhin:2021} and other properties~\cite{cencek01,Puchalski:2016,lesiuk20}, it has been observed that this is frequently the case. The QED effects are only several times smaller than the relativistic corrections in such cases, rather than by a factor close to $\alpha\log\alpha \approx 0.036$.

To account for this phenomena, we settle on the worst case scenario. Namely, instead of scaling the total relativistic correction by a factor of $\alpha\log\alpha$, we chose the relativistic correction which is the largest in magnitude and perform similar scaling. Multiplying the relativistic kinetic-energy correction $\delta\chi_0^{\mathrm{P}_4}$by $\alpha\log\alpha$ we find that the leading QED correction to $\chi_0$ can be roughly estimated as $0.000\,034\times 10^{-5}\,a_0^3$. 
This value most likely overestimates the QED effects, so we view our estimate 
of the uncertainty of $\chi_0$ as rather conservative.

\section{Computational details and numerical results}
\label{sec:kompot}

To evaluate all quantities necessary for the determination of magnetic susceptibility of helium, we follow closely the numerical approach applied in our recent calculations of electric polarizability~\cite{Puchalski:2020}.
The ground-state wavefunction of the helium atom is represented as a linear combination of Slater geminals, namely 
\begin{equation}
\label{basis}
\psi_0(\mathbf{r}_1, \mathbf{r}_2) = (1 + {\cal P}_{12})
\sum_{i=1}^{N} c_i\, e^{-\alpha_i\,r_1 - \beta_i\,r_2 - \gamma_i\,r_{12} },
\end{equation}
where the ${\cal P}_{12}$ operator interchanges coordinates of $\mathbf{r}_1$ and $\mathbf{r}_2$.
The linear coefficients $c_i$ and the nonlinear parameters $\alpha_i$, $\beta_i$, and $\gamma_i$ are fully optimized to minimize the nonrelativistic energy of helium.
For a given set of basis functions, defined by the 
nonlinear parameters $\alpha_i$, $\beta_i$, and $\gamma_i$, the coefficients $c_i$ were obtained using the Rayleigh-Ritz 
method. The standard linear algebra and minimization algorithms implemented in HSL Mathematical Software Library~\cite{HSL} were applied and quadruple precision arithmetics was used to enhance the numerical stability. In particular, matrix factorizations and solutions of corresponding systems of equations were performed employing the DAG-based parallel Cholesky method with OpenMP interface for shared-memory multiprocessing. Full optimization of the nonlinear parameters was carried out by applying two subroutines, i.e. VA13 - the BFGS variable metric method when values of the derivatives with respect to the variables were used and VA24 - the conjugate directions method when these derivatives were not employed. Using two different optimization procedures allowed us to avoid the situation of optimization getting stuck in one of the local minima, and also accelerated the convergence of the whole optimization procedure.
 
The advantage of the exponential basis set~(\ref{basis}) 
is the correct functional form near the interparticle coalescence points (the Kato's cusp), both of the electron-electron and 
electron-nucleus types.
This enables us to determine highly accurate wavefunctions with a relatively compact basis. 
In order to estimate the uncertainty of the results, 
we performed all calculations with a sequence of basis sets 
with $N = 128$, $256$, and $512$ functions. With the largest basis set, the non-relativistic energy is accurate to $17$ significant 
digits as compared to the benchmark value of Ref.~\onlinecite{Aznabaev:2018}. This accuracy guarantees that numerical uncertainties of 
all computed quantities are negligible in comparison with errors resulting from omission of higher-order corrections 
(both in $\alpha$ and $m/m_N$).

\begin{table}[t]
\caption{\label{tab:expect} Expectation values of various operators for the ground state of helium atom computed with the largest basis set of $N=512$ functions. The results are given in atomic units. In the parentheses we show the estimated uncertainty of the last digit. } 
\begin{ruledtabular}
\begin{tabular}{ll}
\phantom{aa} operator & \phantom{aaa} expectation value \\
\hline\\[-1em]
$H_0$   & 
$-$2.903\,724\,377\,034\,119\,59(1) \\
 $r_1^2 + r_2^2 $ &
\phantom{$-$}2.386\,965\,990\,037\,9(1) \\
$ l_1^2 + l_2^2 $ &
\phantom{$-$}0.018\,970\,526\,333(1) \\
$r_1^2\,p_1^2 + r_2^2\,p_2^2 $ & 
$-$0.139\,689\,120\,125(1) \\
$\mathbf{r}_1\!\cdot \mathbf{r}_2 $ & 
$-$0.064\,736\,661\,397\,785(1) \\
${\mathbf{r}_1\!\cdot\mathbf{r}_2}\,\,{r^{-1}_{12}} $ & 
\phantom{$-$}0.059\,280\,414\,991\,545(2) \\
$ {\mathbf{r}_1\!\cdot\mathbf{r}_2}\,\,{r^{-1}_{12}} - 
 {(\mathbf{r}_{12}\!\cdot\mathbf{r}_1)\,{r_{12}^{-3}}(\mathbf{r}_{12}\!\cdot\mathbf{r}_{2})}$ & 
\phantom{$-$}0.212\,506\,954\,000(1) \\
\end{tabular}
\end{ruledtabular}
\end{table}

\begin{table}[t]
\caption{\label{tab:elems}The matrix elements $\langle\psi_0|\big(r_1^2 + r_2^2\big)\mathcal{R}_0X|\psi_0\rangle$ for several operators $X$ required in this work obtained with the largest basis set of $N=512$ functions. The results are given in atomic units. The estimated uncertainty of the last digit is shown in the parentheses.
}
\begin{ruledtabular}
\begin{tabular}{ll}
$X$ & matrix element \\
\hline\\[-1em]
$p_1^4 + p_2^4 $     
& $-$80.298\,613(2) \\
 $ \pi \big[\delta^{3}(\mathbf{r}_1) + \delta^{3}(\mathbf{r}_2)\big] $ 
  & $-$7.918\,414\,9(1) \\
 $ \pi \,\delta^{3}(\mathbf{r}_{12}) $ 
  & $-$0.547\,997\,8(1) \\
$ {\mathbf{p}_1{r^{-1}_{12}} \mathbf{p}_2} + 
 {(\mathbf{p}_1\!\cdot\mathbf{r}_{12}){r^{-3}_{12}}(\mathbf{r}_{12}\cdot\mathbf{p}_2)}$ 
& \phantom{$-$}0.420\,214\,859\,4(2) \\
$\mathbf{p}_1\!\cdot\mathbf{p}_2$ 
& $-$0.179\,805\,762\,988\,59(3) \\
\end{tabular}
\end{ruledtabular}
\end{table}

In Table~\ref{tab:expect} we present expectation values of all operators required to calculate the diamagnetic susceptibility of helium, taking into account the 
finite-nuclear-mass and relativistic corrections considered in Secs.~\ref{sec:fnm}~and~\ref{sec:rel} respectively. The error of each quantity is estimated conservatively as half of the difference between the results obtained with $N=256$ and $N=512$ basis sets.

In order to evaluate the mass-polarization correction~(\ref{fnmmp}), as well as the relativistic Breit-Pauli correction~(\ref{relbp}), one has to compute the following first-order response function
\begin{align}
 \psi_1 = -\mathcal{R}_0\big(r_1^2 + r_2^2\big) \psi_0.
\end{align}
Once the response function $\psi_1$ is known, all these corrections can be rewritten in a form that permits their stable numerical evaluation. In order to obtain $ {\psi}_1$ we first note that it obeys equation 
\begin{align}
 (QH-E_0){\psi}_1 = Q\big(r_1^2 + r_2^2\big) \psi_0
\end{align}
and hence can be found by minimization of the following Hylleraas functional
\begin{align}
\label{hyll}
 \mathcal{F}[\widetilde{\psi}] = \langle \widetilde{\psi}|(H_0-E_0 -E_0P_0)|\widetilde{\psi}\rangle
 + 2\,\langle\widetilde{\psi}|Q\big(r_1^2 + r_2^2\big) | \psi_0\rangle,
\end{align}
where $P_0 =\mid \psi_0\rangle\langle\psi_0|$,
with respect to all parameters appearing in the trial wavefunction $\widetilde{\psi}$.
Since the operator $r_1^2 + r_2^2$ is spherically symmetric 
the trial function $\widetilde{\psi}$ can also be represented by the expansion of the form of Eq.~(\ref{basis}). 
 However, the size of the basis employed in calculation of $\psi_1$ had to be twice as large as employed for $\psi_0$. This basis was generated in the following way. The first part of the basis 
set, comprising $N$ functions, has the same non-linear parameters $\alpha_i$, $\beta_i$, and $\gamma_i$ as found for 
the ground state. This part of the basis is not the subject of further optimization, i.e., the linear parameters 
$\alpha_i$, $\beta_i$, and $\gamma_i$ for $i\leq N$ are fixed and only the expansion coefficients $c_i$ are calculated anew. This 
approach guarantees accurate fulfillment of the orthogonality condition $\langle{\psi}_1|\psi_0\rangle=0$, resulting from the 
presence of the $Q$ projection in the definition of the resolvent. The second part of the basis for 
$\widetilde{\psi}$, also comprising $N$ functions, includes functions with non-linear parameters optimized by minimizing the functional of 
Eq.~(\ref{hyll}). In Table~\ref{tab:elems} we show the numerical results of the second-order matrix elements obtained 
for the ground state of helium atom using the largest basis set $N=512$. The error estimation is performed in the same way 
as for the data given in Table~\ref{tab:expect}.

\section{Discussion and conclusions}
\label{sec:concl}
 
\begin{table*}[ht]
\caption{Contributions to the diamagnetic susceptibility of the $^3$He and $^4$He atoms given in the  units of  $10^{-5}\,a_0^3$.  The numbers in parentheses are the uncertainties of the last digit; when no uncertainty estimate is shown, the last digit is accurate.  The uncertainties of the data from Ref.~\onlinecite{Bruch:2002} were not estimated by the authors and hence are  not shown.}
\label{tab:final}
\begin{ruledtabular}
\begin{tabular}{llll}
contribution & 
\multicolumn{1}{c}{$^4$He} & 
\multicolumn{1}{c}{$^3$He} &
\multicolumn{1}{c}{$^3$He, Refs.~\onlinecite{Bruch:2002,Bruch:2003}} \\
\hline\\[-1.2em]
$\chi_0^{(0)}$ & \multicolumn{2}{c}{$-$2.118\,486\,203\,037\,9(1)} \\
\hline\\[-1.1em]
& \multicolumn{2}{c}{\rm Finite-nuclear-mass correction, $\chi_0^{\mathrm{FNM}}$} \\[0.1em]
 \hline\\[-1.1em]
$\delta\chi_0^{\mathrm{ms}}$ 
& $-$0.000\,871\,291\,146\,3 
&$-$0.001\,156\,403\,069\,7 
& $-$0.001\,16  \\
$\delta\chi_0^{\mathrm{PZW}}$ 
& \phantom{$-$}0.000\,015\,753\,465\,6 
& \phantom{$-$}0.000\,020\,908\,459\,8 
& \phantom{$-$}0.000\,020\,92$^a$ \\
$\delta\chi_0^{\mathrm{mp}}$ 
& $-$0.000\,043\,755\,174\,2 
& $-$0.000\,058\,073\,145\,8 
& $-$0.000\,058\,1 \\
total $\chi_0^{\mathrm{FNM}}$
& $-$0.000\,899\,21
& $-$0.001\,193\,49
& $-$0.001\,20$^b$ \\
\hline\\[-1.1em]
& \multicolumn{2}{c}{\rm Relativistic correction, $\delta\chi_0^{\mathrm{rel}}$} \\
\hline\\[-1.1em]
$\delta\chi_0^{(1)}$ 
& \multicolumn{2}{c}{\phantom{$-$}0.000\,000\,448\,290} 
& \multicolumn{1}{c}{n/a} \\
$\delta\chi_0^{(2)}$ 
& \multicolumn{2}{c}{$-$0.000\,003\,300\,978}
& \multicolumn{1}{c}{n/a} \\
$\delta\chi_0^{(3)}$ 
& \multicolumn{2}{c}{\phantom{$-$}0.000\,141\,785\,338}
& \multicolumn{1}{c}{n/a} \\
$\delta\chi_0^{(4)}$ 
& \multicolumn{2}{c}{\phantom{$-$}0.000\,002\,801\,697}
& \multicolumn{1}{c}{n/a} \\
$\delta\chi_0^{(5)}$ 
&\multicolumn{2}{c}{\phantom{$-$}0.000\,005\,021\,728}
& \multicolumn{1}{c}{n/a} \\
\hline\\[-1.1em]
$\sum_{i}\delta\chi_0^{(i)}$
&\multicolumn{2}{c}{\phantom{$-$}0.000\,146\,756\,076}
& \multicolumn{1}{c}{n/a}\\
\hline\\[-1.1em]
$\delta\chi_0^{\mathrm{P}_4}$ 
& \multicolumn{2}{c}{\phantom{$-$}0.000\,948\,763\,83(2)} 
&\phantom{$-$}0.000\,95 \\
$\delta\chi_0^{\mathrm{D}_1}+\delta\chi_0^{\mathrm{D}_2}$ 
& \multicolumn{2}{c}{$-$0.000\,800\,275\,46(1)} 
& $-$0.000\,802 \\
$\delta\chi_0^{\mathrm{B}}$   
& \multicolumn{2}{c}{\phantom{$-$}0.000\,019\,860\,10\phantom{(0)}} 
& \phantom{$-$} 0.000\,019\,8 \\
$\delta\chi_0^{\mathrm{BP}}$
& \multicolumn{2}{c}{\phantom{$-$}0.000\,168\,348\,47(3)} 
& \phantom{$-$}{0.000\,162\,8  }\\
total $\delta\chi_0^{\mathrm{rel}}$
& \multicolumn{2}{c}{\phantom{$-$}0.000\,315\,104\,55(2)}
& \phantom{$-$}{0.000\,162\,8 } \\[1ex]
\hline\\[-1.1em]
\hline\\[-1.1em]
$\chi_0=\chi_0^{(0)}+\chi_0^{\mathrm{FNM}}+\delta\chi_0^{\mathrm{rel}} $  & $-$2.119\,070(34) & $-$2.119\,365(34) & $-$2.119\,52
\end{tabular}
\end{ruledtabular}
\begin{flushleft}
$^a\,$the original value from Ref.~\onlinecite{Bruch:2002} adjusted by a factor $4$ 
as noted in Ref.~\onlinecite{Bruch:2003}; \\
$^b\,$based on values from Ref.~\onlinecite{Bruch:2002};
\end{flushleft}
\end{table*}  
 
In Table~\ref{tab:final} we present contributions to the diamagnetic susceptibility of the $^3$He and $^4$He atoms based on numerical values from Tables~\ref{tab:expect}~and~\ref{tab:elems}. Our results for $^3$He are compared with the previous results of Bruch and Weinhold from Ref.~\cite {Bruch:2002}. We found a good agreement in all individual contributions computed in Ref.~\cite{Bruch:2002}, including both the $\chi_0^{\mathrm{FNM}}$ and $\delta\chi_0^{\mathrm{BP}}$ corrections. However, the remaining relativistic contributions derived in the present work, namely $\delta\chi_0^{(i)}$, $i=1,\ldots,5$, were not considered in Ref.~\cite{Bruch:2002}. Most of these corrections turned out to be small, with the exception of $\delta\chi_0^{(3)}$ which is of the same order of magnitude as the dominant $\delta\chi_0^{\mathrm{BP}}$ term. Because of that, the total relativistic correction reported in Ref.~\cite {Bruch:2002} is underestimated by a factor of about one third in comparison to our data. Note that in Ref.~\cite {Bruch:2002}, the mass polarization correction $\delta\chi_0^{\mathrm{mp}}$ was treated together with $\delta\chi_0^{\mathrm{BP}}$ rather 
than with $\chi_0^{\mathrm{FNM}}$, as in our work, which would be more appropriate considering the scaling of both terms with the nuclear mass. 
Overall, the present numerical results are of high numerical accuracy. The errors of our calculations, rigorously estimated, are 
negligible in comparison with neglected higher-order order terms in $\alpha$
and in $m_{\rm e}/m_N$. 

In Ref.~\cite {Bruch:2002} Bruch and Weinhold considered also the effect on $\chi_0$ 
due to the the center-of-mass motion, referred to by them as $\delta \chi_0^{\rm BO}$. 
Using two different 
 approximate perturbation theory procedures they derived two order-of-magnitude 
estimations of $\delta \chi_0^{\rm BO}$ that can be expressed by the formula 
\begin{align}
\label{chibo1}
 \delta\chi_0^{\mathrm{BO}}\approx f \frac{E_{\rm cm}}{m_Nc^2}\,\,a_0^3,
\end{align}
where $E_{\rm cm}$ is the center-of-mass kinetic energy of the atom and $f$ is a dimensionless parameter close to 1.5. 
This temperature-dependent correction is proportional to the ratio of the translational energy $E_{\rm cm}=\frac{3}{2} kT $ to the rest mass of the atom and 
turns out to be several orders of magnitude smaller than $\chi_0^{\mathrm{FNM}}$---four orders for the liquid helium temperature of $T$$\approx$20 K and three orders for the average temperature $T$$\approx$296 K of the gas phase measurements \cite{barter60}. 
 Therefore, even the large relative error in the determination of 
$\delta \chi_0^{\rm BO}$ would not be relevant for metrology applications. 

A more serious problem is the omission 
of the QED corrections in the present work. From the discussion in Sec.~\Ref{sec:qed} it is clear that the final 
uncertainty of the magnetic susceptibility comes from the crude estimation of the 
 {QED} correction and our final recommended value of $\chi_0$ for $^4$He, used in the last row 
 Table III, is $0.000\ 034$$\cdot$$10^{-5}\, a_0^3$. 
Further improvements in the accuracy of $\chi_0$ for helium would 
require on more rigorous determination of the QED contributions.

Let us also compare our results with the available experimental data. From the gas phase experiments performed by 
Barter~\emph{et al.}~\cite{barter60} we have the value $\chi_0=-2.26(8)\cdot 10^{-5}$ $a_0^3$ for $^4$He. Therefore, we find a 
roughly $2\sigma$ disagreement with the theoretical result $\chi_0=-2.119\,106(34)\cdot10^{-5}$ $a_0^3$ determined by us. The 
reason for this disagreement is not clear. On the experimental side, another measurement of the magnetic 
susceptibility was performed in the liquid phase for the $^3$He isotope~\cite{thoman96,gould98,mikhalchuk00}. To add to 
the confusion, the experimental results in the liquid and gas phase also do not agree, with the deviation of about 6\%. 
Moreover, the liquid-gas phase discrepancy is inconsistent with theoretical estimates of Bruch and 
Weinhold~\cite{bruch00}, supported by calculations of Komasa~\cite{komasa00}, who argued that the interaction-induced 
increment to the diamagnetic susceptibility of liquid helium is below 1\%.

While the possible sources of error on the experimental side cannot be elucidated in the present 
work, it is                                                               worth discussing the possible sources of the discrepancy resulting from inaccurate calculations or 
incomplete theory. First, we believe that such large deviations cannot be explained by numerical errors or artefacts 
such as basis set incompleteness, etc. in our computations. This is partly due to mature state of technology used for 
accurate calculations for two-electron systems. Additionally, there is a very good agreement between our numerical 
results and the data of Bruch and Weinhold, in all cases where the latter are available. Similarly, we find it extremely 
unlikely that the QED effects bring such large contribution to the magnetic susceptibility. 
This would 
imply a catastrophic failure of the convergence of the QED energy in powers of $\alpha$, 
resulting in the QED effects being about three 
orders of magnitude larger than the $\alpha^2$ relativistic correction. Such phenomenon would be unprecedented and would 
contradict the current knowledge about the accuracy of QED for light system like the helium atom \cite{Yerokhin:2021,Yerokhin:2022}. 

Several other sources of error in theoretical calculations of $\chi_0$ have been discussed, 
such as approximate treatment of the temperature dependence of $\chi_0$ (represented by the $\delta \chi_0^{\rm BO}$ term) or density dependence of $\chi_0$ considered by 
Bruch and Weinhold~\cite{bruch00}. In all cases, these effects cannot explain the observed discrepancy.
It is worth mentioning that the magnetic susceptibility exhibits also a frequency dependence, an effect completely neglected in the present work. However, the frequency dependence of $\chi_0$ originates solely from high-order contributions of the order $1/m_N^2$ or $1/c^6$ and higher, and hence is entirely negligible within the present accuracy requirements.

To conclude, we have reported state-of-the-art theoretical calculations of the static diamagnetic susceptibility of $^3$He 
and $^4$He in the ground $^1S$ electronic state. We have evaluated the complete relativistic correction to 
$\chi_0$ of the order $\alpha^4$, including terms originating from the magnetic-field dependent Dirac equation and the Breit interaction. The correction due to the 
finite nuclear mass has also been evaluated. The main source of error in our calculations is the omission of the QED effects 
which were crudely and very conservatively estimated. Our theoretical results disagree both with the gas-phase and liquid-phase 
measurements of the magnetic susceptibility. The reason for this disagreement is not known; possible sources of error on 
the theoretical side were discussed. A new independent measurement may shed light on this problem and help to resolve 
the discrepancy.

\begin{acknowledgments}
We thank Krzysztof Pachucki for numerous discussions. This project (QuantumPascal project 18SIB04) has received funding from the EMPIR programme cofinanced by the Participating States and from the European Union’s Horizon 2020 research and innovation program. The authors also acknowledge support from the National Science Center, Poland, within the Project No. 2017/27/B/ST4/02739.
\end{acknowledgments}

\appendix

\section{}
\label{sec:app}
Here we give derivation of Eqs.~(\ref{chi4})~and~(\ref{chi5}). Let us first consider $\delta\chi_0^{(4)}$ correction of Eq.~(\Ref{chi4}). Using the vector identity 
\begin{equation}
\label{id}
(\mathbf{a}\times\mathbf{b})\cdot(\mathbf{c}\times\mathbf{d})= 
(\mathbf{a}\cdot \mathbf{c})(\mathbf{b}\cdot \mathbf{d}) - 
(\mathbf{a}\cdot\mathbf{d})(\mathbf{b}\cdot\mathbf{c}), 
\end{equation}
the operator in the numerator in Eq.~(\ref{A44}) can be written as
\begin{align}
\label{a2} 
(\mathbf{B}\times\mathbf{r}_1)\cdot(\mathbf{B}\times\mathbf{r}_2) =
 B^2 \, \mathbf{r}_1\cdot \mathbf{r}_2 - 
 (\mathbf{B}\cdot\mathbf{r}_1)(\mathbf{B}\cdot\mathbf{r}_2).
\end{align}
It is not difficult to see that the expectation value of the second term on the right-hand-side of Eq.~(\ref{a2}), when evaluated with a spherically symmetric wave function, is the same as the expectation value of the operator $-\frac{1}{3} B^2\,\mathbf{r}_1\! \cdot\mathbf{r}_2$. This allows us to write:
\begin{align}
 \langle\psi_0|\frac{(\mathbf{B}\times\mathbf{r}_1) \cdot(\mathbf{B}\times\mathbf{r}_2)}{r_{12}}|\psi_0\rangle= 
 \frac{2}{3}B^2\,
 \langle\psi_0|\frac{\mathbf{r}_1\!\cdot \mathbf{r}_2}{r_{12}}|\psi_0\rangle.
\end{align}
Double differentiation with respect to $B$ generates an additional factor of $2$ which finally leads to Eq.~(\ref{chi4}).

Derivation of Eq.~(\ref{chi5}) is somewat more complicated. First, by expanding the vector and scalar products appearing in Eq.~(\ref{d6}) we obtain:
\begin{align} 
\begin{split}
 & \big[(\mathbf{B}\!\times\mathbf{r}_1)\cdot \mathbf{r}_{12} \big]
 \big[(\mathbf{B}\times\mathbf{r}_2)\cdot\mathbf{r}_{12} \big] = \\
 & \big[ \mathbf{B}\cdot(\mathbf{r}_1\!\times \mathbf{r}_{12}) \big]
 \big[ \mathbf{B}\cdot(\mathbf{r}_2\!\times\mathbf{r}_{12} )\big] 
 \\ \label{A4}
 &= \frac{1}{3} B^2 \, (\mathbf{r}_1\!\times \mathbf{r}_{12}) \cdot
 (\mathbf{r}_2\times\mathbf{r}_{12} ) + \ldots ,
\end{split}
\end{align}
where the dots indicate several terms that give zero when evaluated with a spherically symmetric wave function. Using Eq.(~\Ref{id}) again one obtains 
\begin{align}
\begin{split}
 &\frac{1}{3}B^2\langle\psi_0|\frac{\, (\mathbf{r}_1\!\times \mathbf{r}_{12}) \cdot
 (\mathbf{r}_2\times\mathbf{r}_{12} )}{r_{12}^3}|\psi_0\rangle= \\
 &\frac{1}{3}B^2
 \langle\psi_0|\frac{r_{12}^2(\mathbf{r}_1\!\cdot \mathbf{r}_2)
 -(\mathbf{r}_1\!\cdot\mathbf{r}_{12})(\mathbf{r}_2\!\cdot\mathbf{r}_{12})}{r_{12}^3}
 |\psi_0\rangle. \label{A5}
\end{split}
\end{align} 
 Differentiation with respect to the external magnetic field leads to Eq.~(\ref{chi5}).
Note that the term explicitly written on the rightmost in Eq.~(\ref{A4}) can also be expressed as $\frac{1}{3}B^2 \, (\mathbf{r}_1\times \mathbf{r}_{ 2})\cdot (\mathbf{r}_1\times\mathbf{r}_{ 2} )$. Thus, in view of Eq.~(\Ref{id}), the right-hand side of Eq.~(\Ref{A5}) can be written in a formally somewhat simpler form 
\begin{equation}
 \frac{1}{3}B^2\langle\psi_0|\frac{r_{1}^2r_{2}^2 
 -(\mathbf{r}_1\!\cdot\mathbf{r}_{2})^2 }{r_{12}^3}
 |\psi_0\rangle. 
\end{equation}

\bibliography{chi_he}

\begin{thebibliography}{44}%
\makeatletter
\providecommand \@ifxundefined [1]{%
 \@ifx{#1\undefined}
}%
\providecommand \@ifnum [1]{%
 \ifnum #1\expandafter \@firstoftwo
 \else \expandafter \@secondoftwo
 \fi
}%
\providecommand \@ifx [1]{%
 \ifx #1\expandafter \@firstoftwo
 \else \expandafter \@secondoftwo
 \fi
}%
\providecommand \natexlab [1]{#1}%
\providecommand \enquote  [1]{``#1''}%
\providecommand \bibnamefont  [1]{#1}%
\providecommand \bibfnamefont [1]{#1}%
\providecommand \citenamefont [1]{#1}%
\providecommand \href@noop [0]{\@secondoftwo}%
\providecommand \href [0]{\begingroup \@sanitize@url \@href}%
\providecommand \@href[1]{\@@startlink{#1}\@@href}%
\providecommand \@@href[1]{\endgroup#1\@@endlink}%
\providecommand \@sanitize@url [0]{\catcode `\\12\catcode `\$12\catcode
  `\&12\catcode `\#12\catcode `\^12\catcode `\_12\catcode `\%12\relax}%
\providecommand \@@startlink[1]{}%
\providecommand \@@endlink[0]{}%
\providecommand \url  [0]{\begingroup\@sanitize@url \@url }%
\providecommand \@url [1]{\endgroup\@href {#1}{\urlprefix }}%
\providecommand \urlprefix  [0]{URL }%
\providecommand \Eprint [0]{\href }%
\providecommand \doibase [0]{http://dx.doi.org/}%
\providecommand \selectlanguage [0]{\@gobble}%
\providecommand \bibinfo  [0]{\@secondoftwo}%
\providecommand \bibfield  [0]{\@secondoftwo}%
\providecommand \translation [1]{[#1]}%
\providecommand \BibitemOpen [0]{}%
\providecommand \bibitemStop [0]{}%
\providecommand \bibitemNoStop [0]{.\EOS\space}%
\providecommand \EOS [0]{\spacefactor3000\relax}%
\providecommand \BibitemShut  [1]{\csname bibitem#1\endcsname}%
\let\auto@bib@innerbib\@empty
\bibitem [{\citenamefont {Yerokhin}\ \emph {et~al.}(2011)\citenamefont
  {Yerokhin}, \citenamefont {Pachucki}, \citenamefont {Harman},\ and\
  \citenamefont {Keitel}}]{yerokhin11}%
  \BibitemOpen
  \bibfield  {author} {\bibinfo {author} {\bibfnamefont {V.~A.}\ \bibnamefont
  {Yerokhin}}, \bibinfo {author} {\bibfnamefont {K.}~\bibnamefont {Pachucki}},
  \bibinfo {author} {\bibfnamefont {Z.}~\bibnamefont {Harman}}, \ and\ \bibinfo
  {author} {\bibfnamefont {C.~H.}\ \bibnamefont {Keitel}},\ }\href@noop {}
  {\bibfield  {journal} {\bibinfo  {journal} {Phys. Rev. Lett.}\ }\textbf
  {\bibinfo {volume} {107}},\ \bibinfo {pages} {043004} (\bibinfo {year}
  {2011})}\BibitemShut {NoStop}%
\bibitem [{\citenamefont {Yerokhin}\ \emph {et~al.}(2012)\citenamefont
  {Yerokhin}, \citenamefont {Pachucki}, \citenamefont {Harman},\ and\
  \citenamefont {Keitel}}]{yerokhin12}%
  \BibitemOpen
  \bibfield  {author} {\bibinfo {author} {\bibfnamefont {V.~A.}\ \bibnamefont
  {Yerokhin}}, \bibinfo {author} {\bibfnamefont {K.}~\bibnamefont {Pachucki}},
  \bibinfo {author} {\bibfnamefont {Z.}~\bibnamefont {Harman}}, \ and\ \bibinfo
  {author} {\bibfnamefont {C.~H.}\ \bibnamefont {Keitel}},\ }\href@noop {}
  {\bibfield  {journal} {\bibinfo  {journal} {Phys. Rev. A}\ }\textbf {\bibinfo
  {volume} {85}},\ \bibinfo {pages} {022512} (\bibinfo {year}
  {2012})}\BibitemShut {NoStop}%
\bibitem [{\citenamefont {Lesiuk}\ \emph {et~al.}(2020)\citenamefont {Lesiuk},
  \citenamefont {Przybytek},\ and\ \citenamefont {Jeziorski}}]{lesiuk20}%
  \BibitemOpen
  \bibfield  {author} {\bibinfo {author} {\bibfnamefont {M.}~\bibnamefont
  {Lesiuk}}, \bibinfo {author} {\bibfnamefont {M.}~\bibnamefont {Przybytek}}, \
  and\ \bibinfo {author} {\bibfnamefont {B.}~\bibnamefont {Jeziorski}},\
  }\href@noop {} {\bibfield  {journal} {\bibinfo  {journal} {Physical Review
  A}\ }\textbf {\bibinfo {volume} {102}},\ \bibinfo {pages} {052816} (\bibinfo
  {year} {2020})}\BibitemShut {NoStop}%
\bibitem [{\citenamefont {Jousten}\ \emph {et~al.}(2017)\citenamefont
  {Jousten}, \citenamefont {Hendricks}, \citenamefont {Barker}, \citenamefont
  {Douglas}, \citenamefont {Eckel}, \citenamefont {Egan}, \citenamefont
  {Fedchak}, \citenamefont {Fl{\"u}gge}, \citenamefont {Gaiser}, \citenamefont
  {Olson} \emph {et~al.}}]{jousten17}%
  \BibitemOpen
  \bibfield  {author} {\bibinfo {author} {\bibfnamefont {K.}~\bibnamefont
  {Jousten}}, \bibinfo {author} {\bibfnamefont {J.}~\bibnamefont {Hendricks}},
  \bibinfo {author} {\bibfnamefont {D.}~\bibnamefont {Barker}}, \bibinfo
  {author} {\bibfnamefont {K.}~\bibnamefont {Douglas}}, \bibinfo {author}
  {\bibfnamefont {S.}~\bibnamefont {Eckel}}, \bibinfo {author} {\bibfnamefont
  {P.}~\bibnamefont {Egan}}, \bibinfo {author} {\bibfnamefont {J.}~\bibnamefont
  {Fedchak}}, \bibinfo {author} {\bibfnamefont {J.}~\bibnamefont {Fl{\"u}gge}},
  \bibinfo {author} {\bibfnamefont {C.}~\bibnamefont {Gaiser}}, \bibinfo
  {author} {\bibfnamefont {D.}~\bibnamefont {Olson}},  \emph {et~al.},\
  }\href@noop {} {\bibfield  {journal} {\bibinfo  {journal} {Metrologia}\
  }\textbf {\bibinfo {volume} {54}},\ \bibinfo {pages} {S146} (\bibinfo {year}
  {2017})}\BibitemShut {NoStop}%
\bibitem [{\citenamefont {Gaiser}\ and\ \citenamefont
  {Fellmuth}(2018)}]{gaiser18}%
  \BibitemOpen
  \bibfield  {author} {\bibinfo {author} {\bibfnamefont {C.}~\bibnamefont
  {Gaiser}}\ and\ \bibinfo {author} {\bibfnamefont {B.}~\bibnamefont
  {Fellmuth}},\ }\href@noop {} {\bibfield  {journal} {\bibinfo  {journal}
  {Phys. Rev. Lett.}\ }\textbf {\bibinfo {volume} {120}},\ \bibinfo {pages}
  {123203} (\bibinfo {year} {2018})}\BibitemShut {NoStop}%
\bibitem [{\citenamefont {Gaiser}\ \emph {et~al.}(2020)\citenamefont {Gaiser},
  \citenamefont {Fellmuth},\ and\ \citenamefont {Sabuga}}]{gaiser20}%
  \BibitemOpen
  \bibfield  {author} {\bibinfo {author} {\bibfnamefont {C.}~\bibnamefont
  {Gaiser}}, \bibinfo {author} {\bibfnamefont {B.}~\bibnamefont {Fellmuth}}, \
  and\ \bibinfo {author} {\bibfnamefont {W.}~\bibnamefont {Sabuga}},\
  }\href@noop {} {\bibfield  {journal} {\bibinfo  {journal} {Nat. Phys.}\
  }\textbf {\bibinfo {volume} {16}},\ \bibinfo {pages} {177} (\bibinfo {year}
  {2020})}\BibitemShut {NoStop}%
\bibitem [{\citenamefont {Gaiser}\ \emph {et~al.}(2022)\citenamefont {Gaiser},
  \citenamefont {Fellmuth},\ and\ \citenamefont {Sabuga}}]{gaiser22}%
  \BibitemOpen
  \bibfield  {author} {\bibinfo {author} {\bibfnamefont {C.}~\bibnamefont
  {Gaiser}}, \bibinfo {author} {\bibfnamefont {B.}~\bibnamefont {Fellmuth}}, \
  and\ \bibinfo {author} {\bibfnamefont {W.}~\bibnamefont {Sabuga}},\
  }\href@noop {} {\bibfield  {journal} {\bibinfo  {journal} {Ann. Phys.}\
  }\textbf {\bibinfo {volume} {534}},\ \bibinfo {pages} {2200336} (\bibinfo
  {year} {2022})}\BibitemShut {NoStop}%
\bibitem [{\citenamefont {Gao}\ \emph {et~al.}(2017)\citenamefont {Gao},
  \citenamefont {Pitre}, \citenamefont {Luo}, \citenamefont {Plimmer},
  \citenamefont {Lin}, \citenamefont {Zhang}, \citenamefont {Feng},
  \citenamefont {Chen},\ and\ \citenamefont {Sparasci}}]{gao17}%
  \BibitemOpen
  \bibfield  {author} {\bibinfo {author} {\bibfnamefont {B.}~\bibnamefont
  {Gao}}, \bibinfo {author} {\bibfnamefont {L.}~\bibnamefont {Pitre}}, \bibinfo
  {author} {\bibfnamefont {E.}~\bibnamefont {Luo}}, \bibinfo {author}
  {\bibfnamefont {M.}~\bibnamefont {Plimmer}}, \bibinfo {author} {\bibfnamefont
  {P.}~\bibnamefont {Lin}}, \bibinfo {author} {\bibfnamefont {J.}~\bibnamefont
  {Zhang}}, \bibinfo {author} {\bibfnamefont {X.}~\bibnamefont {Feng}},
  \bibinfo {author} {\bibfnamefont {Y.}~\bibnamefont {Chen}}, \ and\ \bibinfo
  {author} {\bibfnamefont {F.}~\bibnamefont {Sparasci}},\ }\href@noop {}
  {\bibfield  {journal} {\bibinfo  {journal} {Measurement}\ }\textbf {\bibinfo
  {volume} {103}},\ \bibinfo {pages} {258} (\bibinfo {year}
  {2017})}\BibitemShut {NoStop}%
\bibitem [{\citenamefont {Rourke}\ \emph {et~al.}(2019)\citenamefont {Rourke},
  \citenamefont {Gaiser}, \citenamefont {Gao}, \citenamefont {Ripa},
  \citenamefont {Moldover}, \citenamefont {Pitre},\ and\ \citenamefont
  {Underwood}}]{rourke19}%
  \BibitemOpen
  \bibfield  {author} {\bibinfo {author} {\bibfnamefont {P.~M.}\ \bibnamefont
  {Rourke}}, \bibinfo {author} {\bibfnamefont {C.}~\bibnamefont {Gaiser}},
  \bibinfo {author} {\bibfnamefont {B.}~\bibnamefont {Gao}}, \bibinfo {author}
  {\bibfnamefont {D.~M.}\ \bibnamefont {Ripa}}, \bibinfo {author}
  {\bibfnamefont {M.~R.}\ \bibnamefont {Moldover}}, \bibinfo {author}
  {\bibfnamefont {L.}~\bibnamefont {Pitre}}, \ and\ \bibinfo {author}
  {\bibfnamefont {R.~J.}\ \bibnamefont {Underwood}},\ }\href@noop {} {\bibfield
   {journal} {\bibinfo  {journal} {Metrologia}\ }\textbf {\bibinfo {volume}
  {56}},\ \bibinfo {pages} {032001} (\bibinfo {year} {2019})}\BibitemShut
  {NoStop}%
\bibitem [{\citenamefont {Ripa}\ \emph {et~al.}(2021)\citenamefont {Ripa},
  \citenamefont {Imbraguglio}, \citenamefont {Gaiser}, \citenamefont {Steur},
  \citenamefont {Giraudi}, \citenamefont {Fogliati}, \citenamefont
  {Bertinetti}, \citenamefont {Lopardo}, \citenamefont {Dematteis},\ and\
  \citenamefont {Gavioso}}]{ripa21}%
  \BibitemOpen
  \bibfield  {author} {\bibinfo {author} {\bibfnamefont {D.~M.}\ \bibnamefont
  {Ripa}}, \bibinfo {author} {\bibfnamefont {D.}~\bibnamefont {Imbraguglio}},
  \bibinfo {author} {\bibfnamefont {C.}~\bibnamefont {Gaiser}}, \bibinfo
  {author} {\bibfnamefont {P.}~\bibnamefont {Steur}}, \bibinfo {author}
  {\bibfnamefont {D.}~\bibnamefont {Giraudi}}, \bibinfo {author} {\bibfnamefont
  {M.}~\bibnamefont {Fogliati}}, \bibinfo {author} {\bibfnamefont
  {M.}~\bibnamefont {Bertinetti}}, \bibinfo {author} {\bibfnamefont
  {G.}~\bibnamefont {Lopardo}}, \bibinfo {author} {\bibfnamefont
  {R.}~\bibnamefont {Dematteis}}, \ and\ \bibinfo {author} {\bibfnamefont
  {R.}~\bibnamefont {Gavioso}},\ }\href@noop {} {\bibfield  {journal} {\bibinfo
   {journal} {Metrologia}\ }\textbf {\bibinfo {volume} {58}},\ \bibinfo {pages}
  {025008} (\bibinfo {year} {2021})}\BibitemShut {NoStop}%
\bibitem [{\citenamefont {Rourke}(2021)}]{rourke21}%
  \BibitemOpen
  \bibfield  {author} {\bibinfo {author} {\bibfnamefont {P.~M.}\ \bibnamefont
  {Rourke}},\ }\href@noop {} {\bibfield  {journal} {\bibinfo  {journal} {J.
  Phys. Chem. Ref. Data}\ }\textbf {\bibinfo {volume} {50}},\ \bibinfo {pages}
  {033104} (\bibinfo {year} {2021})}\BibitemShut {NoStop}%
\bibitem [{\citenamefont {Lorentz}(1880)}]{lorentz80}%
  \BibitemOpen
  \bibfield  {author} {\bibinfo {author} {\bibfnamefont {H.~A.}\ \bibnamefont
  {Lorentz}},\ }\href@noop {} {\bibfield  {journal} {\bibinfo  {journal} {Ann.
  Phys.}\ }\textbf {\bibinfo {volume} {245}},\ \bibinfo {pages} {641} (\bibinfo
  {year} {1880})}\BibitemShut {NoStop}%
\bibitem [{\citenamefont {Lorenz}(1880)}]{lorenz80}%
  \BibitemOpen
  \bibfield  {author} {\bibinfo {author} {\bibfnamefont {L.}~\bibnamefont
  {Lorenz}},\ }\href@noop {} {\bibfield  {journal} {\bibinfo  {journal} {Ann.
  Phys.}\ }\textbf {\bibinfo {volume} {247}},\ \bibinfo {pages} {70} (\bibinfo
  {year} {1880})}\BibitemShut {NoStop}%
\bibitem [{\citenamefont {Pachucki}\ and\ \citenamefont
  {Sapirstein}(2000)}]{pachucki00}%
  \BibitemOpen
  \bibfield  {author} {\bibinfo {author} {\bibfnamefont {K.}~\bibnamefont
  {Pachucki}}\ and\ \bibinfo {author} {\bibfnamefont {J.}~\bibnamefont
  {Sapirstein}},\ }\href@noop {} {\bibfield  {journal} {\bibinfo  {journal}
  {Phys. Rev. A}\ }\textbf {\bibinfo {volume} {63}},\ \bibinfo {pages} {012504}
  (\bibinfo {year} {2000})}\BibitemShut {NoStop}%
\bibitem [{\citenamefont {\L{}ach}\ \emph {et~al.}(2004)\citenamefont
  {\L{}ach}, \citenamefont {Jeziorski},\ and\ \citenamefont
  {Szalewicz}}]{lach04}%
  \BibitemOpen
  \bibfield  {author} {\bibinfo {author} {\bibfnamefont {G.}~\bibnamefont
  {\L{}ach}}, \bibinfo {author} {\bibfnamefont {B.}~\bibnamefont {Jeziorski}},
  \ and\ \bibinfo {author} {\bibfnamefont {K.}~\bibnamefont {Szalewicz}},\
  }\href@noop {} {\bibfield  {journal} {\bibinfo  {journal} {Phys. Rev. Lett.}\
  }\textbf {\bibinfo {volume} {92}},\ \bibinfo {pages} {233001} (\bibinfo
  {year} {2004})}\BibitemShut {NoStop}%
\bibitem [{\citenamefont {Piszczatowski}\ \emph {et~al.}(2015)\citenamefont
  {Piszczatowski}, \citenamefont {Puchalski}, \citenamefont {Komasa},
  \citenamefont {Jeziorski},\ and\ \citenamefont
  {Szalewicz}}]{Piszczatowski:2015}%
  \BibitemOpen
  \bibfield  {author} {\bibinfo {author} {\bibfnamefont {K.}~\bibnamefont
  {Piszczatowski}}, \bibinfo {author} {\bibfnamefont {M.}~\bibnamefont
  {Puchalski}}, \bibinfo {author} {\bibfnamefont {J.}~\bibnamefont {Komasa}},
  \bibinfo {author} {\bibfnamefont {B.}~\bibnamefont {Jeziorski}}, \ and\
  \bibinfo {author} {\bibfnamefont {K.}~\bibnamefont {Szalewicz}},\ }\href@noop
  {} {\bibfield  {journal} {\bibinfo  {journal} {Phys. Rev. Lett.}\ }\textbf
  {\bibinfo {volume} {114}},\ \bibinfo {pages} {173004} (\bibinfo {year}
  {2015})}\BibitemShut {NoStop}%
\bibitem [{\citenamefont {Puchalski}\ \emph {et~al.}(2016)\citenamefont
  {Puchalski}, \citenamefont {Piszczatowski}, \citenamefont {Komasa},
  \citenamefont {Jeziorski},\ and\ \citenamefont {Szalewicz}}]{Puchalski:2016}%
  \BibitemOpen
  \bibfield  {author} {\bibinfo {author} {\bibfnamefont {M.}~\bibnamefont
  {Puchalski}}, \bibinfo {author} {\bibfnamefont {K.}~\bibnamefont
  {Piszczatowski}}, \bibinfo {author} {\bibfnamefont {J.}~\bibnamefont
  {Komasa}}, \bibinfo {author} {\bibfnamefont {B.}~\bibnamefont {Jeziorski}}, \
  and\ \bibinfo {author} {\bibfnamefont {K.}~\bibnamefont {Szalewicz}},\
  }\href@noop {} {\bibfield  {journal} {\bibinfo  {journal} {Phys. Rev. A}\
  }\textbf {\bibinfo {volume} {93}},\ \bibinfo {pages} {032515} (\bibinfo
  {year} {2016})}\BibitemShut {NoStop}%
\bibitem [{\citenamefont {Puchalski}\ \emph
  {et~al.}(2020{\natexlab{a}})\citenamefont {Puchalski}, \citenamefont
  {Szalewicz}, \citenamefont {Lesiuk},\ and\ \citenamefont
  {Jeziorski}}]{puchalski20}%
  \BibitemOpen
  \bibfield  {author} {\bibinfo {author} {\bibfnamefont {M.}~\bibnamefont
  {Puchalski}}, \bibinfo {author} {\bibfnamefont {K.}~\bibnamefont
  {Szalewicz}}, \bibinfo {author} {\bibfnamefont {M.}~\bibnamefont {Lesiuk}}, \
  and\ \bibinfo {author} {\bibfnamefont {B.}~\bibnamefont {Jeziorski}},\
  }\href@noop {} {\bibfield  {journal} {\bibinfo  {journal} {Phys. Rev. A}\
  }\textbf {\bibinfo {volume} {101}},\ \bibinfo {pages} {022505} (\bibinfo
  {year} {2020}{\natexlab{a}})}\BibitemShut {NoStop}%
\bibitem [{\citenamefont {Bruch}\ and\ \citenamefont
  {Weinhold}(2002)}]{Bruch:2002}%
  \BibitemOpen
  \bibfield  {author} {\bibinfo {author} {\bibfnamefont {L.~W.}\ \bibnamefont
  {Bruch}}\ and\ \bibinfo {author} {\bibfnamefont {F.}~\bibnamefont
  {Weinhold}},\ }\href@noop {} {\bibfield  {journal} {\bibinfo  {journal} {J.
  Chem. Phys.}\ }\textbf {\bibinfo {volume} {117}},\ \bibinfo {pages} {3243}
  (\bibinfo {year} {2002})}\BibitemShut {NoStop}%
\bibitem [{\citenamefont {Bruch}\ and\ \citenamefont
  {Weinhold}(2003)}]{Bruch:2003}%
  \BibitemOpen
  \bibfield  {author} {\bibinfo {author} {\bibfnamefont {L.~W.}\ \bibnamefont
  {Bruch}}\ and\ \bibinfo {author} {\bibfnamefont {F.}~\bibnamefont
  {Weinhold}},\ }\href@noop {} {\bibfield  {journal} {\bibinfo  {journal} {J.
  Chem. Phys.}\ }\textbf {\bibinfo {volume} {119}},\ \bibinfo {pages} {638}
  (\bibinfo {year} {2003})}\BibitemShut {NoStop}%
\bibitem [{\citenamefont {Barter}\ \emph {et~al.}(1960)\citenamefont {Barter},
  \citenamefont {Meisenheimer},\ and\ \citenamefont {Stevenson}}]{barter60}%
  \BibitemOpen
  \bibfield  {author} {\bibinfo {author} {\bibfnamefont {C.}~\bibnamefont
  {Barter}}, \bibinfo {author} {\bibfnamefont {R.}~\bibnamefont
  {Meisenheimer}}, \ and\ \bibinfo {author} {\bibfnamefont {D.}~\bibnamefont
  {Stevenson}},\ }\href@noop {} {\bibfield  {journal} {\bibinfo  {journal} {J.
  Phys. Chem.}\ }\textbf {\bibinfo {volume} {64}},\ \bibinfo {pages} {1312}
  (\bibinfo {year} {1960})}\BibitemShut {NoStop}%
\bibitem [{\citenamefont {Pachucki}(2003)}]{pachucki03}%
  \BibitemOpen
  \bibfield  {author} {\bibinfo {author} {\bibfnamefont {K.}~\bibnamefont
  {Pachucki}},\ }\href@noop {} {\bibfield  {journal} {\bibinfo  {journal}
  {Phys. Rev. A}\ }\textbf {\bibinfo {volume} {67}},\ \bibinfo {pages} {012504}
  (\bibinfo {year} {2003})}\BibitemShut {NoStop}%
\bibitem [{\citenamefont {Bethe}\ and\ \citenamefont
  {Salpeter}(1977)}]{Bethe:77}%
  \BibitemOpen
  \bibfield  {author} {\bibinfo {author} {\bibfnamefont {H.~A.}\ \bibnamefont
  {Bethe}}\ and\ \bibinfo {author} {\bibfnamefont {E.~E.}\ \bibnamefont
  {Salpeter}},\ }\href@noop {} {\emph {\bibinfo {title} {Quantum mechanics of
  one- and two-electron atoms}}}\ (\bibinfo  {publisher} {Plenum},\ \bibinfo
  {address} {New York},\ \bibinfo {year} {1977})\BibitemShut {NoStop}%
\bibitem [{\citenamefont {Landau}\ and\ \citenamefont
  {Lifshitz}(1981)}]{landau81}%
  \BibitemOpen
  \bibfield  {author} {\bibinfo {author} {\bibfnamefont {L.~D.}\ \bibnamefont
  {Landau}}\ and\ \bibinfo {author} {\bibfnamefont {L.~M.}\ \bibnamefont
  {Lifshitz}},\ }\href@noop {} {\emph {\bibinfo {title} {Quantum Mechanics
  Non-Relativistic Theory, Third Edition: Volume 3}}},\ \bibinfo {edition}
  {3rd}\ ed.\ (\bibinfo  {publisher} {Butterworth-Heinemann},\ \bibinfo {year}
  {1981})\BibitemShut {NoStop}%
\bibitem [{\citenamefont {Pachucki}\ and\ \citenamefont
  {Yerokhin}(2019)}]{Pachucki:2019}%
  \BibitemOpen
  \bibfield  {author} {\bibinfo {author} {\bibfnamefont {K.}~\bibnamefont
  {Pachucki}}\ and\ \bibinfo {author} {\bibfnamefont {V.~A.}\ \bibnamefont
  {Yerokhin}},\ }\href@noop {} {\bibfield  {journal} {\bibinfo  {journal}
  {Phys. Rev. A}\ }\textbf {\bibinfo {volume} {100}},\ \bibinfo {pages}
  {062510} (\bibinfo {year} {2019})}\BibitemShut {NoStop}%
\bibitem [{\citenamefont {Pachucki}(2008)}]{Pachucki:2008}%
  \BibitemOpen
  \bibfield  {author} {\bibinfo {author} {\bibfnamefont {K.}~\bibnamefont
  {Pachucki}},\ }\href@noop {} {\bibfield  {journal} {\bibinfo  {journal}
  {Phys. Rev. A}\ }\textbf {\bibinfo {volume} {78}},\ \bibinfo {pages} {012504}
  (\bibinfo {year} {2008})}\BibitemShut {NoStop}%
\bibitem [{\citenamefont {Wehrli}\ \emph {et~al.}(2021)\citenamefont {Wehrli},
  \citenamefont {Spyszkiewicz-Kaczmarek}, \citenamefont {Puchalski},\ and\
  \citenamefont {Pachucki}}]{Wehrli:2021}%
  \BibitemOpen
  \bibfield  {author} {\bibinfo {author} {\bibfnamefont {D.}~\bibnamefont
  {Wehrli}}, \bibinfo {author} {\bibfnamefont {A.}~\bibnamefont
  {Spyszkiewicz-Kaczmarek}}, \bibinfo {author} {\bibfnamefont {M.}~\bibnamefont
  {Puchalski}}, \ and\ \bibinfo {author} {\bibfnamefont {K.}~\bibnamefont
  {Pachucki}},\ }\href@noop {} {\bibfield  {journal} {\bibinfo  {journal}
  {Phys. Rev. Lett.}\ }\textbf {\bibinfo {volume} {127}},\ \bibinfo {pages}
  {263001} (\bibinfo {year} {2021})}\BibitemShut {NoStop}%
\bibitem [{\citenamefont {Wehrli}\ \emph {et~al.}(2022)\citenamefont {Wehrli},
  \citenamefont {Puchalski},\ and\ \citenamefont {Pachucki}}]{Wehrli:2022}%
  \BibitemOpen
  \bibfield  {author} {\bibinfo {author} {\bibfnamefont {D.}~\bibnamefont
  {Wehrli}}, \bibinfo {author} {\bibfnamefont {M.}~\bibnamefont {Puchalski}}, \
  and\ \bibinfo {author} {\bibfnamefont {K.}~\bibnamefont {Pachucki}},\
  }\href@noop {} {\bibfield  {journal} {\bibinfo  {journal} {Phys. Rev. A}\
  }\textbf {\bibinfo {volume} {105}},\ \bibinfo {pages} {032808} (\bibinfo
  {year} {2022})}\BibitemShut {NoStop}%
\bibitem [{\citenamefont {Pachucki}(2006)}]{pachucki06}%
  \BibitemOpen
  \bibfield  {author} {\bibinfo {author} {\bibfnamefont {K.}~\bibnamefont
  {Pachucki}},\ }\href@noop {} {\bibfield  {journal} {\bibinfo  {journal}
  {Phys. Rev. A}\ }\textbf {\bibinfo {volume} {74}},\ \bibinfo {pages} {022512}
  (\bibinfo {year} {2006})}\BibitemShut {NoStop}%
\bibitem [{\citenamefont {Piszczatowski}\ \emph {et~al.}(2009)\citenamefont
  {Piszczatowski}, \citenamefont {{\L}ach}, \citenamefont {Przybytek},
  \citenamefont {Komasa}, \citenamefont {Pachucki},\ and\ \citenamefont
  {Jeziorski}}]{piszcz09}%
  \BibitemOpen
  \bibfield  {author} {\bibinfo {author} {\bibfnamefont {K.}~\bibnamefont
  {Piszczatowski}}, \bibinfo {author} {\bibfnamefont {G.}~\bibnamefont
  {{\L}ach}}, \bibinfo {author} {\bibfnamefont {M.}~\bibnamefont {Przybytek}},
  \bibinfo {author} {\bibfnamefont {J.}~\bibnamefont {Komasa}}, \bibinfo
  {author} {\bibfnamefont {K.}~\bibnamefont {Pachucki}}, \ and\ \bibinfo
  {author} {\bibfnamefont {B.}~\bibnamefont {Jeziorski}},\ }\href@noop {}
  {\bibfield  {journal} {\bibinfo  {journal} {J. Chem. Theory Comput.}\
  }\textbf {\bibinfo {volume} {5}},\ \bibinfo {pages} {3039} (\bibinfo {year}
  {2009})}\BibitemShut {NoStop}%
\bibitem [{\citenamefont {Cencek}\ \emph {et~al.}(2012)\citenamefont {Cencek},
  \citenamefont {Przybytek}, \citenamefont {Komasa}, \citenamefont {Mehl},
  \citenamefont {Jeziorski},\ and\ \citenamefont {Szalewicz}}]{cencek12}%
  \BibitemOpen
  \bibfield  {author} {\bibinfo {author} {\bibfnamefont {W.}~\bibnamefont
  {Cencek}}, \bibinfo {author} {\bibfnamefont {M.}~\bibnamefont {Przybytek}},
  \bibinfo {author} {\bibfnamefont {J.}~\bibnamefont {Komasa}}, \bibinfo
  {author} {\bibfnamefont {J.~B.}\ \bibnamefont {Mehl}}, \bibinfo {author}
  {\bibfnamefont {B.}~\bibnamefont {Jeziorski}}, \ and\ \bibinfo {author}
  {\bibfnamefont {K.}~\bibnamefont {Szalewicz}},\ }\href@noop {} {\bibfield
  {journal} {\bibinfo  {journal} {J. Chem. Phys.}\ }\textbf {\bibinfo {volume}
  {136}},\ \bibinfo {pages} {224303} (\bibinfo {year} {2012})}\BibitemShut
  {NoStop}%
\bibitem [{\citenamefont {Lesiuk}\ \emph {et~al.}(2015)\citenamefont {Lesiuk},
  \citenamefont {Przybytek}, \citenamefont {Musia\l{}}, \citenamefont
  {Jeziorski},\ and\ \citenamefont {Moszynski}}]{lesiuk15}%
  \BibitemOpen
  \bibfield  {author} {\bibinfo {author} {\bibfnamefont {M.}~\bibnamefont
  {Lesiuk}}, \bibinfo {author} {\bibfnamefont {M.}~\bibnamefont {Przybytek}},
  \bibinfo {author} {\bibfnamefont {M.}~\bibnamefont {Musia\l{}}}, \bibinfo
  {author} {\bibfnamefont {B.}~\bibnamefont {Jeziorski}}, \ and\ \bibinfo
  {author} {\bibfnamefont {R.}~\bibnamefont {Moszynski}},\ }\href@noop {}
  {\bibfield  {journal} {\bibinfo  {journal} {Phys. Rev. A}\ }\textbf {\bibinfo
  {volume} {91}},\ \bibinfo {pages} {012510} (\bibinfo {year}
  {2015})}\BibitemShut {NoStop}%
\bibitem [{\citenamefont {Lesiuk}\ \emph {et~al.}(2019)\citenamefont {Lesiuk},
  \citenamefont {Przybytek}, \citenamefont {Balcerzak}, \citenamefont
  {Musia{\l}},\ and\ \citenamefont {Moszynski}}]{lesiuk19}%
  \BibitemOpen
  \bibfield  {author} {\bibinfo {author} {\bibfnamefont {M.}~\bibnamefont
  {Lesiuk}}, \bibinfo {author} {\bibfnamefont {M.}~\bibnamefont {Przybytek}},
  \bibinfo {author} {\bibfnamefont {J.~G.}\ \bibnamefont {Balcerzak}}, \bibinfo
  {author} {\bibfnamefont {M.}~\bibnamefont {Musia{\l}}}, \ and\ \bibinfo
  {author} {\bibfnamefont {R.}~\bibnamefont {Moszynski}},\ }\href@noop {}
  {\bibfield  {journal} {\bibinfo  {journal} {J. Chem. Theory Comput.}\
  }\textbf {\bibinfo {volume} {15}},\ \bibinfo {pages} {2470} (\bibinfo {year}
  {2019})}\BibitemShut {NoStop}%
\bibitem [{\citenamefont {Yerokhin}\ \emph
  {et~al.}(2021{\natexlab{a}})\citenamefont {Yerokhin}, \citenamefont
  {Patko\v{s}},\ and\ \citenamefont {Pachucki}}]{Yerokhin:2021}%
  \BibitemOpen
  \bibfield  {author} {\bibinfo {author} {\bibfnamefont {V.~A.}\ \bibnamefont
  {Yerokhin}}, \bibinfo {author} {\bibfnamefont {V.}~\bibnamefont
  {Patko\v{s}}}, \ and\ \bibinfo {author} {\bibfnamefont {K.}~\bibnamefont
  {Pachucki}},\ }\href@noop {} {\bibfield  {journal} {\bibinfo  {journal}
  {Symmetry}\ }\textbf {\bibinfo {volume} {13}},\ \bibinfo {pages} {1246}
  (\bibinfo {year} {2021}{\natexlab{a}})}\BibitemShut {NoStop}%
\bibitem [{\citenamefont {Cencek}\ \emph {et~al.}(2001)\citenamefont {Cencek},
  \citenamefont {Szalewicz},\ and\ \citenamefont {Jeziorski}}]{cencek01}%
  \BibitemOpen
  \bibfield  {author} {\bibinfo {author} {\bibfnamefont {W.}~\bibnamefont
  {Cencek}}, \bibinfo {author} {\bibfnamefont {K.}~\bibnamefont {Szalewicz}}, \
  and\ \bibinfo {author} {\bibfnamefont {B.}~\bibnamefont {Jeziorski}},\
  }\href@noop {} {\bibfield  {journal} {\bibinfo  {journal} {Phys. Rev. Lett.}\
  }\textbf {\bibinfo {volume} {86}},\ \bibinfo {pages} {5675} (\bibinfo {year}
  {2001})}\BibitemShut {NoStop}%
\bibitem [{\citenamefont {Puchalski}\ \emph
  {et~al.}(2020{\natexlab{b}})\citenamefont {Puchalski}, \citenamefont
  {Szalewicz}, \citenamefont {Lesiuk},\ and\ \citenamefont
  {Jeziorski}}]{Puchalski:2020}%
  \BibitemOpen
  \bibfield  {author} {\bibinfo {author} {\bibfnamefont {M.}~\bibnamefont
  {Puchalski}}, \bibinfo {author} {\bibfnamefont {K.}~\bibnamefont
  {Szalewicz}}, \bibinfo {author} {\bibfnamefont {M.}~\bibnamefont {Lesiuk}}, \
  and\ \bibinfo {author} {\bibfnamefont {B.}~\bibnamefont {Jeziorski}},\
  }\href@noop {} {\bibfield  {journal} {\bibinfo  {journal} {Phys. Rev. A}\
  }\textbf {\bibinfo {volume} {101}},\ \bibinfo {pages} {022505} (\bibinfo
  {year} {2020}{\natexlab{b}})}\BibitemShut {NoStop}%
\bibitem [{HSL()}]{HSL}%
  \BibitemOpen
  \href@noop {} {\bibinfo  {journal} {HSL. A collection of Fortran codes for
  large scale scientific computation. http://www.hsl.rl.ac.uk/}\ }\BibitemShut
  {NoStop}%
\bibitem [{\citenamefont {Aznabaev}\ \emph {et~al.}(2018)\citenamefont
  {Aznabaev}, \citenamefont {Bekbaev},\ and\ \citenamefont
  {Korobov}}]{Aznabaev:2018}%
  \BibitemOpen
\bibfield  {journal} {  }\bibfield  {author} {\bibinfo {author} {\bibfnamefont
  {D.~T.}\ \bibnamefont {Aznabaev}}, \bibinfo {author} {\bibfnamefont {A.~K.}\
  \bibnamefont {Bekbaev}}, \ and\ \bibinfo {author} {\bibfnamefont {V.~I.}\
  \bibnamefont {Korobov}},\ }\href@noop {} {\bibfield  {journal} {\bibinfo
  {journal} {Phys. Rev. A}\ }\textbf {\bibinfo {volume} {98}},\ \bibinfo
  {pages} {012510} (\bibinfo {year} {2018})}\BibitemShut {NoStop}%
\bibitem [{\citenamefont {Thoman}\ \emph {et~al.}(1996)\citenamefont {Thoman},
  \citenamefont {Mikhalchuk}, \citenamefont {Bozler},\ and\ \citenamefont
  {Gould}}]{thoman96}%
  \BibitemOpen
  \bibfield  {author} {\bibinfo {author} {\bibfnamefont {M.}~\bibnamefont
  {Thoman}}, \bibinfo {author} {\bibfnamefont {A.}~\bibnamefont {Mikhalchuk}},
  \bibinfo {author} {\bibfnamefont {H.}~\bibnamefont {Bozler}}, \ and\ \bibinfo
  {author} {\bibfnamefont {C.}~\bibnamefont {Gould}},\ }\href@noop {}
  {\bibfield  {journal} {\bibinfo  {journal} {Czechoslov. J. Phys.}\ }\textbf
  {\bibinfo {volume} {46}},\ \bibinfo {pages} {229} (\bibinfo {year}
  {1996})}\BibitemShut {NoStop}%
\bibitem [{\citenamefont {Gould}\ and\ \citenamefont {Bozler}(1998)}]{gould98}%
  \BibitemOpen
  \bibfield  {author} {\bibinfo {author} {\bibfnamefont {C.}~\bibnamefont
  {Gould}}\ and\ \bibinfo {author} {\bibfnamefont {H.}~\bibnamefont {Bozler}},\
  }\href@noop {} {\bibfield  {journal} {\bibinfo  {journal} {J. Low Temp.
  Phys.}\ }\textbf {\bibinfo {volume} {113}},\ \bibinfo {pages} {661} (\bibinfo
  {year} {1998})}\BibitemShut {NoStop}%
\bibitem [{\citenamefont {Mikhalchuk}\ \emph {et~al.}(2000)\citenamefont
  {Mikhalchuk}, \citenamefont {White}, \citenamefont {Bozler},\ and\
  \citenamefont {Gould}}]{mikhalchuk00}%
  \BibitemOpen
  \bibfield  {author} {\bibinfo {author} {\bibfnamefont {A.}~\bibnamefont
  {Mikhalchuk}}, \bibinfo {author} {\bibfnamefont {K.}~\bibnamefont {White}},
  \bibinfo {author} {\bibfnamefont {H.}~\bibnamefont {Bozler}}, \ and\ \bibinfo
  {author} {\bibfnamefont {C.}~\bibnamefont {Gould}},\ }\href@noop {}
  {\bibfield  {journal} {\bibinfo  {journal} {Phys. B: Condens. Matter}\
  }\textbf {\bibinfo {volume} {284}},\ \bibinfo {pages} {238} (\bibinfo {year}
  {2000})}\BibitemShut {NoStop}%
\bibitem [{\citenamefont {Bruch}\ and\ \citenamefont
  {Weinhold}(2000)}]{bruch00}%
  \BibitemOpen
  \bibfield  {author} {\bibinfo {author} {\bibfnamefont {L.}~\bibnamefont
  {Bruch}}\ and\ \bibinfo {author} {\bibfnamefont {F.}~\bibnamefont
  {Weinhold}},\ }\href@noop {} {\bibfield  {journal} {\bibinfo  {journal} {J.
  Chem. Phys.}\ }\textbf {\bibinfo {volume} {113}},\ \bibinfo {pages} {8667}
  (\bibinfo {year} {2000})}\BibitemShut {NoStop}%
\bibitem [{\citenamefont {Komasa}(2000)}]{komasa00}%
  \BibitemOpen
  \bibfield  {author} {\bibinfo {author} {\bibfnamefont {J.}~\bibnamefont
  {Komasa}},\ }\href@noop {} {\bibfield  {journal} {\bibinfo  {journal} {J.
  Chem. Phys.}\ }\textbf {\bibinfo {volume} {112}},\ \bibinfo {pages} {7075}
  (\bibinfo {year} {2000})}\BibitemShut {NoStop}%
\bibitem [{\citenamefont {Yerokhin}\ \emph
  {et~al.}(2021{\natexlab{b}})\citenamefont {Yerokhin}, \citenamefont
  {Patko\v{s}},\ and\ \citenamefont {Pachucki}}]{Yerokhin:2022}%
  \BibitemOpen
  \bibfield  {author} {\bibinfo {author} {\bibfnamefont {V.~A.}\ \bibnamefont
  {Yerokhin}}, \bibinfo {author} {\bibfnamefont {V.}~\bibnamefont
  {Patko\v{s}}}, \ and\ \bibinfo {author} {\bibfnamefont {K.}~\bibnamefont
  {Pachucki}},\ }\href@noop {} {\bibfield  {journal} {\bibinfo  {journal}
  {PRA}\ }\textbf {\bibinfo {volume} {106}},\ \bibinfo {pages} {022815}
  (\bibinfo {year} {2021}{\natexlab{b}})}\BibitemShut {NoStop}%
\end{thebibliography}%

\end{document}